# Atmospheric Pressure Mass Spectrometry by Single-Mode Nanoelectromechanical Systems

Batuhan E. Kaynak[1,2,‡], Mohammed Alkhaled[1,2,‡], Enise Kartal[1,2], Cenk Yanik[3], M. Selim Hanay[1,2,*]

[1] Department of Mechanical Engineering, Bilkent University, 06800 Ankara Turkey

[2] UNAM — Institute of Materials Science and Nanotechnology, Bilkent University, 06800 Ankara Turkey

[3] SUNUM, Sabancı University Nanotechnology Research and Application Center, 34956 Istanbul Turkey

[‡] These authors contributed equally.

[*] Corresponding author: selimhanay@bilkent.edu.tr

**ABSTRACT**

Weighing particles above MegaDalton mass range has been a persistent challenge in commercial mass spectrometry. Recently, nanoelectromechanical systems-based mass spectrometry (NEMS-MS) has shown remarkable performance in this mass range, especially with the advance of performing mass spectrometry under entirely atmospheric conditions. This advance reduces the overall complexity and cost, while improving the limit of detection. However, this technique required the tracking of two mechanical modes, and the accurate knowledge of mode shapes which may deviate from their ideal values especially due to air damping. Here, we used a NEMS architecture with a central platform, which enables the calculation of mass by single mode measurements. Experiments were conducted using polystyrene and gold nanoparticles to demonstrate the successful acquisition of mass spectra using a single mode, with improved areal capture efficiency. This advance represents a step forward in NEMS-MS, bringing it closer to becoming a practical application for mass sensing of nanoparticles.

**Keywords**

**Introduction**

Nanoelectromechanical Systems (NEMS) have proven their use in the mass spectrometry field[1-19] for almost two decades, especially for analytes with masses that are unreachable by conventional mass spectrometry techniques, *i.e.* >10 MDa due to large mass-to-charge ratios (m/z).[20-22] Therefore, the ability to measure these high mass values allows NEMS Mass Spectrometry (NEMS-MS) to be a potent tool for the characterization of metallic, ceramic, polymeric, and biological nanoparticles, *e.g.* exosomes, viruses, and lipid vesicles. Our recent study[16] enabled the NEMS-MS technique to work under entirely atmospheric conditions, thus opening possibilities to enhance the NEMS-MS technique while resolving major problems such as low capture efficiencies and high system cost. This technique does not suffer from bulky vacuum elements leading to a lower system cost, and increases the capture efficiencies when compared with the systems that are deployed inside vacuum systems owing to the implementation of the polymeric focusing lens.

Typical device architectures in the NEMS-MS field have been doubly-clamped beams, or cantilevers,[23, 24] with the notable exception of the use of membranes for higher collection efficiencies.[17, 19] In devices with beam architectures, two or more mechanical modes are needed to resolve the landing position and mass of each particle. Apart from requiring two measurement channels running in parallel, the applicability of multimode techniques requires the knowledge of mode shapes which deviate from their ideal values due to non-idealities in nanofabrication, the local stress on the device, and random accumulation of adsorbates on the surface. Besides these factors, a recent study[25] uncovered that the displacement profiles of the NEMS devices are altered to some degree when the NEMS is operated under atmospheric conditions due to the higher dissipation[26-30] (*i.e.,* lower quality factor) caused by viscous damping of air compared to the vacuum, when the devices are driven from one extremity of the device.[31-34] This difference is evident in flexural resonance modes, which are needed for the accurate reverse calculation of the mass and the landing position of an analyte particle. Therefore, the governing equations for multimodal detection now introduce uncertainties for obtaining the mass spectrum of the particles. This alteration in the displacement profiles creates the need for devices that will directly relate the frequency shifts caused by a landing event to the mass of the device while being independent of the mechanical mode shapes in the governing equations. Indeed, motivated in part by these considerations, recent studies with optomechanical sensors under vacuum conditions utilized a platform device with a uniform displacement field to measure, and hence a single mechanical mode shape of the device was used.[15]

In this work, we have investigated the performance of a different class of NEMS architecture —paddle NEMS device— for NEMS-MS under atmospheric conditions. The paddle NEMS device has a platform section in the active region to exhibit a uniform mode shape compared to the conventional architectures reported in the literature, such as doubly-clamped beams. This uniform mode shape enables us to calculate the mass of the landing analytes on the platform independent of the mode shape; rather frequency shifts can be directly related to the mass of the device (for particles that land on the central paddle region). Therefore, the need for the use of higher mechanical modes and reverse-calculation equations that depend on mode shapes — which can cause further uncertainties for atmospheric operation— is eliminated. Moreover, the central placement and large area of the platform increase the unit-area collection efficiency by four times, when compared with the doubly-clamped beams.

In the following, we first detected fluorescent nanoparticles using the paddle NEMS device and investigated these measurements by observing the same particles under fluorescent microscopy. In this case, we electrosprayed 200 nm fluorescent polystyrene nanoparticles (F-PSNP) on the NEMS device and after collecting several such events, we classified the location of the nanoparticles, and related their locations with the observed frequency shifts. This way we validated that for events landing on the central platform, the response of the NEMS is almost uniform.

After the validation of the mass sensing experiments with uniform mode shape, we tested the performance of the paddle NEMS device in sensing by spraying a large number of 40 nm gold nanoparticles (GNP). In this run, we observed a peak in the inferred diameter histogram that corresponds to a mean value of 42.6 nm, which again shows that the system is capable of measuring nanoparticles using only one mechanical mode. Therefore, the paddle NEMS devices decrease the electronic requirement to a single PLL circuitry while eliminating the uncertainties originating from the deviation of the displacement profiles from their ideal values (*i.e.* mode shapes).

**Results/Discussion**

**Paddle Nanoelectromechanical Devices**

Paddle NEMS devices are suspended structures made of LPCVD-grown, low-stress silicon nitride on a silicon substrate (Figure 1a). The fabrication of the device consists of electron beam lithography and photolithography techniques, along with the dry etching steps to suspend the device. The supporting beam structures are typically 4 µm long and 400 nm wide on both sides, and the platform is 3 µm wide and 2 µm long. The fabrication flow is the same with a doubly-clamped beam,[35, 36] including the integration step of the photoresist window.[16]    The implementation of the photoresist window facilitates the delivery of ions to the device surface (Figure 1a,b.). The actuation and detection electrodes are u-shaped resistances that enable thermoelastic actuation and piezoresistive detection (Figure 1b).[35-37]

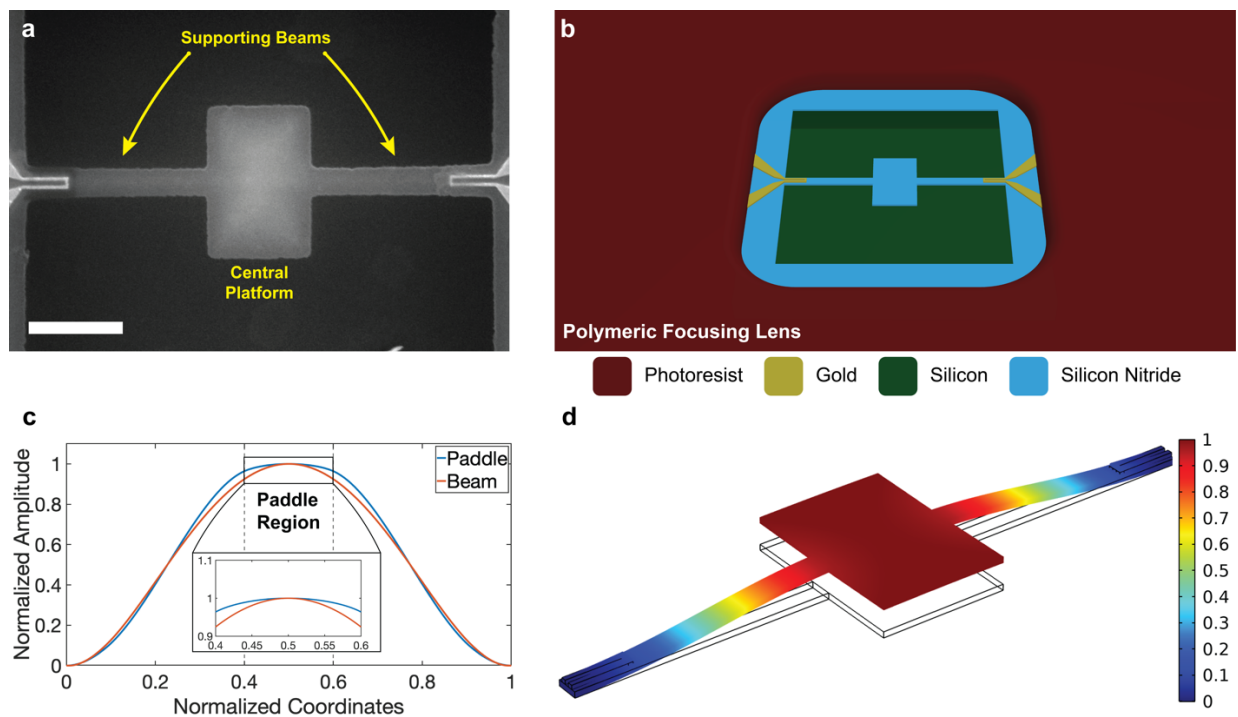

**Figure 1.** Device architecture and mode shape simulations. (a) SEM picture featuring a paddle NEMS device with central platform and supporting beams to connect the device to the anchor points. Scale bar is 2 µm. (b) Render of the device illustrating different layers on the substrate. (c) Mode shape of the paddle NEMS device compared to a normal doubly-clamped beam. (d) First out-of-plane mode shape ($f$ =5.57 MHz) of the paddle NEMS device. Colormap corresponds to the normalized displacement field.

In the typical doubly-clamped beam architecture, the mode shape of the fundamental mechanical mode resembles a half-wavelength sine wave with an anti-node at the middle of the device (Figure 1c). For the paddle NEMS device, numerical simulations indicate that the central platform has a uniform mode shape with a maximum of 5.2% difference between the extremums (Figure 1c,d). Over the entire platform, the standard deviation of the normalized displacement is ~1% (Figure 1d). The use of platform not only makes the mode shape uniform across a large area of the sensor but also increases the total sensor area. Indeed, the central platform comprises a large portion of the active sensor —up to 65% of the sensor's total area. This large area in the middle of the structure is the main region intended to detect the nanoparticles. However, the supporting beams on both sides of the platform are also responsive to the particles adsorbing on them; this situation currently introduces uncertainty in the mass determination for the proposed technique.

Since the frequency shift caused by a particle is a function of the position in the case of doubly-clamped beams, it is necessary to track the first two out-of-plane mechanical modes to resolve the landing position and the mass of the particle. However, in paddle NEMS devices, owing to the uniform mode shape on the platform, the frequency shift is a function of the device mass only, assuming the particle lands on the platform. Accordingly, we only tracked the first mechanical mode and observed the performance of the device in this setting.

**Uniform Mode-Shape Validation Experiments with Fluorescent Nanoparticles**

We first conducted experiments with 200 nm F-PSNP to validate the uniformity of the mode shape on the platform of the paddle NEMS devices. The NEMS chip was placed in front of a custom-made ESI setup which generates ionized nanoparticles which are then transported to the NEMS sensor with the help of the on-chip ion lens. In this experiment, we tracked the first mechanical mode since this mode features a uniform displacement profile across the middle platform, with an Allan Deviation of $4.03 \times 10^{-6}$ at the PLL time scale (set at 140 ms). The Allan Deviation is only three times worse when compared to the doubly-clamped beams operated in atmospheric conditions despite the increase in surface area (see Supporting Figure S1). The minimum detectable mass of this device is 22.7 ag (13.7 MDa) under atmospheric conditions.

In three consecutive sections, a total of 8 particle landing events were detected in the PLL of the first mechanical mode (Figure 2a). The number of events was intentionally kept low to be able to relate the NEMS events to the microscopy images[38] (attaining low-event rates was facilitated by having two devices in one large focusing window). From the simulations, the particles landing at the platform, are expected to induce frequency shifts of similar magnitudes. We indeed observed very similar values for frequency shifts for the six out of eight landing events. The microscope image of the tracked device can be seen in Figure 2b, and the fluorescent microscope image of the device after the last experiment can be seen in Figure 2c. From the fluorescent microscope image in Figure 2c, it is clear that, indeed there are eight particles collected on the device, with six of them on the platform region (there are also additional particles close to the clamping points, but they generate much smaller frequency shifts). Out of the six events, one of them is located right at the edge intersection of the platform with the supporting beam. On Table 1, we listed these 6 landing events that are similar in terms of frequency shifts.

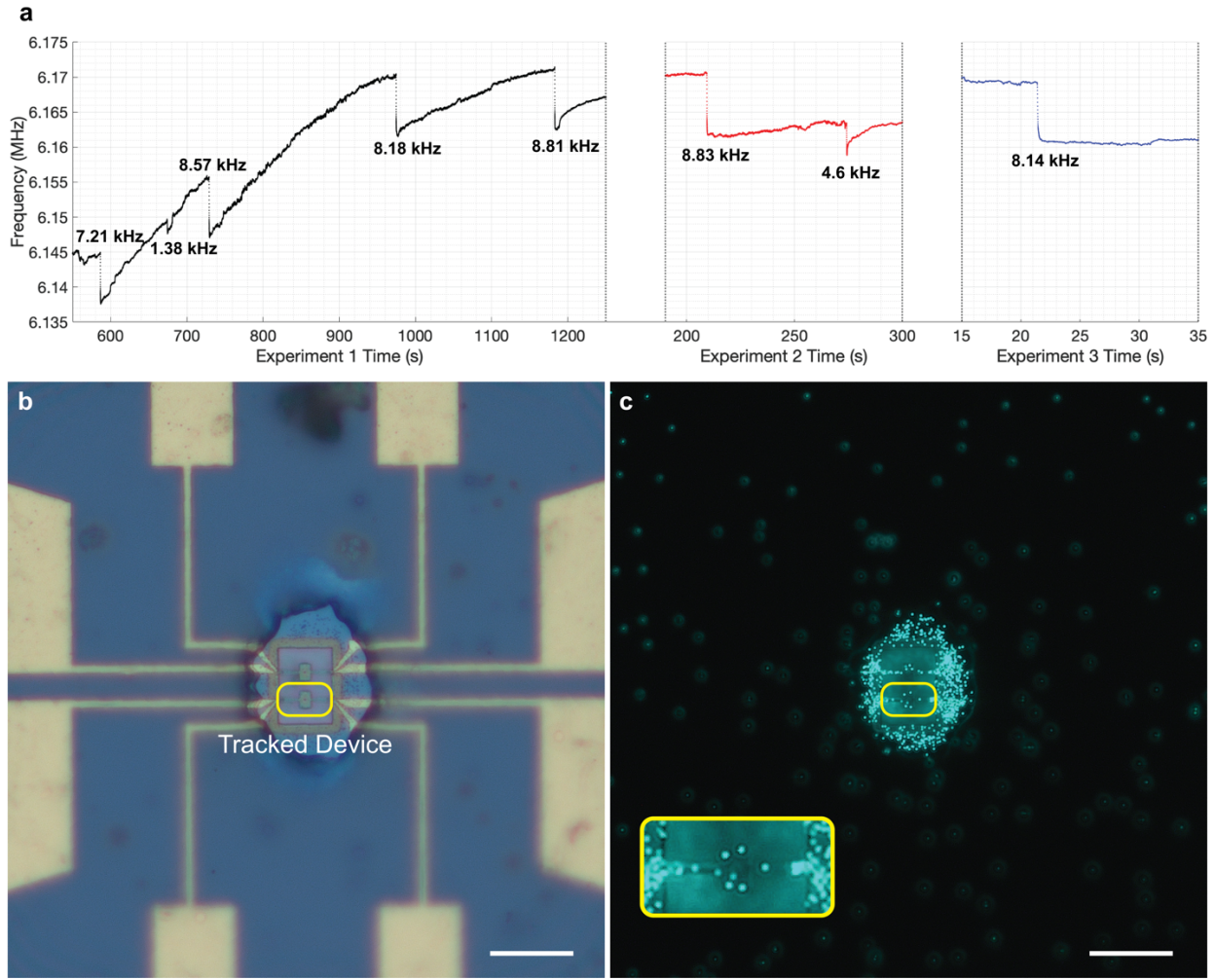

**Figure 2.** Uniform mode shape validation experiments for paddle NEMS devices. (a) PLL data of three consecutive experiments showing eight landing events on the tracked device. (b) Microscope image of the device after the third experiment. The yellow frame indicates the tracked device. Scale bar is 15 μm. (c) Fluorescence image of the device after the third experiment, showing 6 single 200 nm fluorescent polystyrene nanoparticles on the platform of the device (where the mode shape is uniform) and 2 on the supporting beams. The focusing capability of the device is evident by comparing the density of particles inside the window versus those outside. Scale bar is 15 μm. Inset shows the zoomed-in image of the tracked device.

**Table 1.** A detailed analysis of the landing events in Figure 2a. [*] The coefficient of variation is 3.9% if the last event with significantly lower frequency is omitted: this event is deemed to occur at the edge of the platform.

| Events | 1 | 2 | 3 | 4 | 5 | 6 |
|---|---|---|---|---|---|---|
| Frequency Shifts (Hz) | 8140 | 8567 | 8176 | 8808 | 8831 | 7212 |
| Inferred Diameter (nm) | 212.1 | 215.9 | 212.4 | 217.7 | 217.9 | 204.0 |
| Mean (Hz) | 8289 | | | | | |
| Std. Deviation (Hz) | 606 | | | | | |
| Coef. of Variation[*] | 7.3% | | | | | |

The landing events that are on the platform have a mean frequency shift of 8289 Hz with a standard deviation of 606 Hz. The coefficient of variation was calculated as 7.3% for the six landing events, which is smaller than the reported value of the polydispersity of the nanoparticles by the vendor (reported as <10% for diameter). In agreement with the NEMS measurements, we can see in Figure 2c that six of the particles came to different locations on the platform, strongly indicating that these six particles correspond to the set of six frequency shifts with similar values in Table 1. We also observed that one of the frequency shifts (the last entry in Table 1) is smaller in frequency shift compared to the other five. We attribute this frequency shift to the particle on the left edge of the platform, where the displacement is smaller (Figure 2c, inset). If we were to exclude this event from statistics, we can calculate a coefficient of variation of only 3.9%. The frequency shifts in the experiments, the fluorescent images, and the low coefficient of variation between events further validate that the mode shape is uniform on the platform for paddle NEMS devices. The coefficient of variation observed in Table 1 (7.3%) is larger than the coefficient of variation expected from the variation of responsivity (i.e. the square of the mode shape) on the platform (2.2%). We attribute this difference to the inherent polydispersity of the F-PSNPs (<10% for diameter). Thus, any landing event on the platform with the same-sized particles creates frequency shifts with similar dispersion levels. The remaining two frequency shifts, with smaller values of 1.38 kHz and 4.6 kHz, can be attributed to the two particles that can be seen on each side of the supporting beams (Figure 2c inset).

**Mass Spectrum of 200 nm Polystyrene Particles**

In the next set of experiments, we increased the number of 200 nm fluorescent polystyrene nanoparticles collected on the paddle NEMS device by fabricating a NEMS chip with a single NEMS inside the focusing window which was narrower and encapsulated the device more tightly. The Allan Deviation was calculated as $2.75 \times 10^{-6}$ (see Supporting Figure S2) for this device (Figure 3d) which corresponds to a minimum detectable mass of 20 ag (12 MDa) for particles landing on the platform (see Supporting Information). This value is comparable to the doubly-clamped beam resonators used in the literature under atmospheric conditions.

In the experiment, the paddle NEMS device captured 29 landing events – again to ensure that NEMS measurements can be related to microscopy images. A section of the PLL where there are 6 landing events can be seen in Figure 3a. From these 29 landing events, we constructed the frequency shift histogram, which can be seen in Figure 3b. We observed a peak at the expected frequency shift value of 200 nm F-PSNP in the histogram (the mean value of the frequency shifts is $8.55\times10^{-4}$). Also, we observed more events in the lower frequency shift range compared to the higher frequency shifts, and we attribute this to the particles that landed on the supporting beams. The particles landing on the supporting beams are expected to create lower frequency shifts when compared to the particles on the platform due to the lower displacement, thereby introducing uncertainty for the final mass spectrum (see Figure 1d). After acquiring the frequency shift values, we converted these values to mass values (see Supporting Information) first, and then, by assuming particles as perfect spheres and using the density of polystyrene ($1.05$ g/cm$^3$) we converted the mass values to obtain a histogram of inferred diameter values (Figure 3c). In the inferred diameter histogram (Figure 3c), we observed a peak at the expected 200 nm level, with almost half the spectrum (fourteen events) accumulated at or adjacent to the 200 nm histogram bin (since the two adjacent bins are still within the specified range reported by the vendor).

After the experiment, we took both brightfield and fluorescent images of the device, which can be seen in Figure 3d and Figure 3e, respectively. In the fluorescent image, we observed thirteen particles on the platform, which correlates with the number of events in the inferred diameter histogram with fourteen particles having diameters close to 200 nm (Figure 3c). Also, the histogram resulted in an average diameter of 195.5 nm, which is close to the expected value. Therefore, the paddle NEMS devices allowed us to demonstrate the validation of the uniform mode shape on the platform by experimental means. The results show that paddle NEMS devices are able to perform mode shape independent mass sensing experiments under atmospheric conditions.

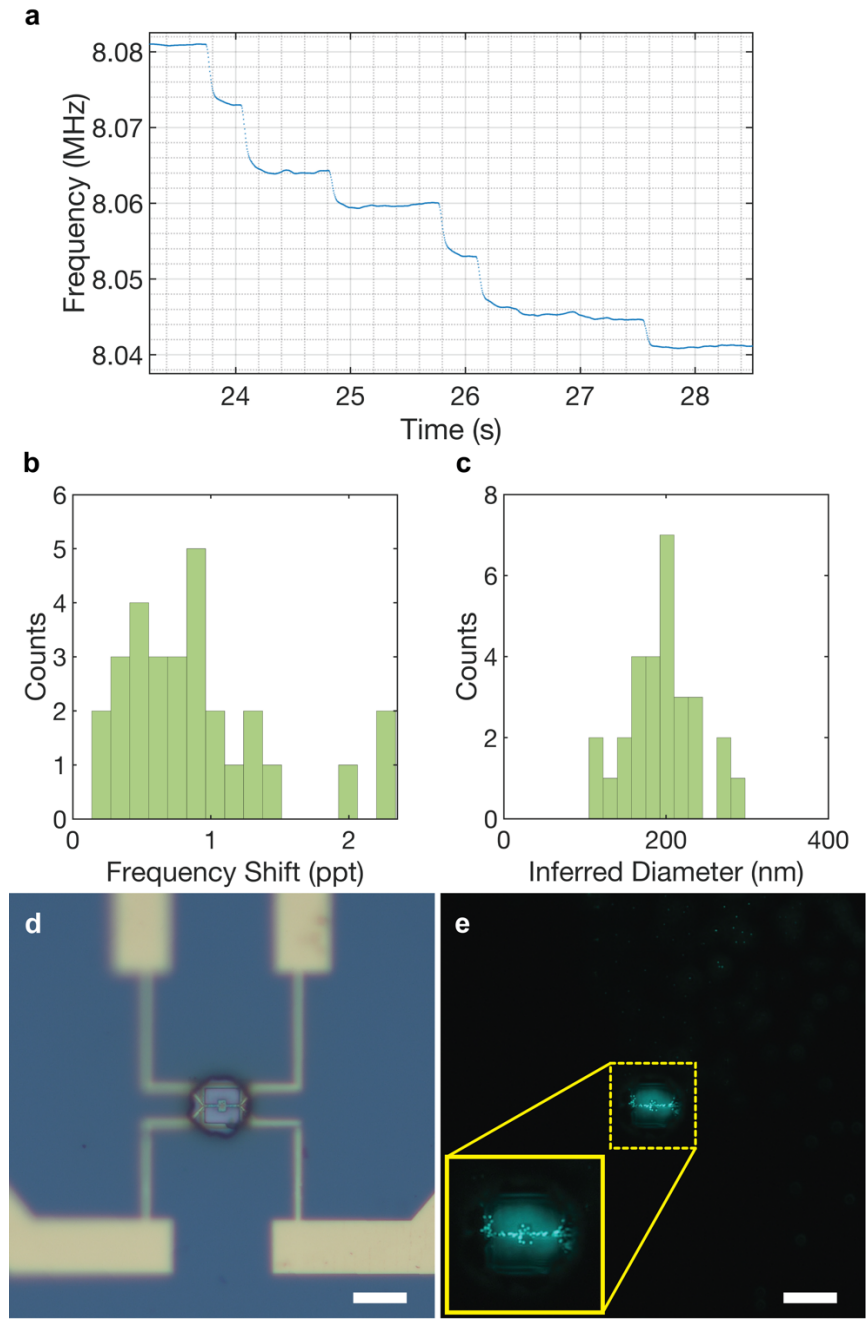

**Figure 3.** Paddle NEMS device used for mass spectrum experiments of 200 nm F-PSNP. (a) A section of the PLL data shows 6 consequent landing events. (b) Frequency shift histogram. (c) Inferred diameter histograms constructed from all 29 landing events. Bin size is 17.5 nm (d) Brightfield and (e) fluorescent images of the device after the experiment. Scale bars are 15 μm for both images. Inset in (e) shows the zoomed in fluorescent image of the device where the single particles on the middle platform can be observed.

The device in Figure 3 contains one paddle NEMS device inside a lensing window. The sensitive paddle area is located at the center of this window, and its large area can collect particles more efficiently than a doubly-clamped beam, as expected of two-dimensional geometry.[17, 19] The capture efficiency per unit area was observed to improve by a factor of four times compared to the earlier polystyrene experiments with doubly-clamped beams (SI table 1).[16] In the calculation of capture efficiency, only the particles landing on the central platform were included (*i.e.* 14 events).

**40 nm Gold Nanoparticle Sensing Experiments**

After validating the uniform mode shape on the platform for paddle NEMS devices, we tested a 40 nm gold nanoparticle sample (Nanopartz A11-40). After taking the open loop sweep, we configured the PLLs and calculated the Allan Deviation as $6.35 \times 10^{-6}$ (See Supporting Figure S5) and the minimum detectable mass values as 40.3 ag (24.3 MDa).

Next, we electrosprayed the 40 nm GNP solution onto the NEMS chip (with a concentration of $8.53 \times 10^{9}$ particles/mL). A section of the PLL data can be seen in Figure 4a. Throughout the measurement, 139 events were collected. In the frequency shift histogram (Figure 4b), where we observed a peak around the expected normalized frequency shift for a 40 nm GNP $(1.44 \times 10^{-4})$ which we attribute to the particle events that we have detected on the platform. The histogram's lower end suggests particles landing on the supporting beams, which results in shifts lower than the expected mean value, due to the smaller displacement at the landing position.

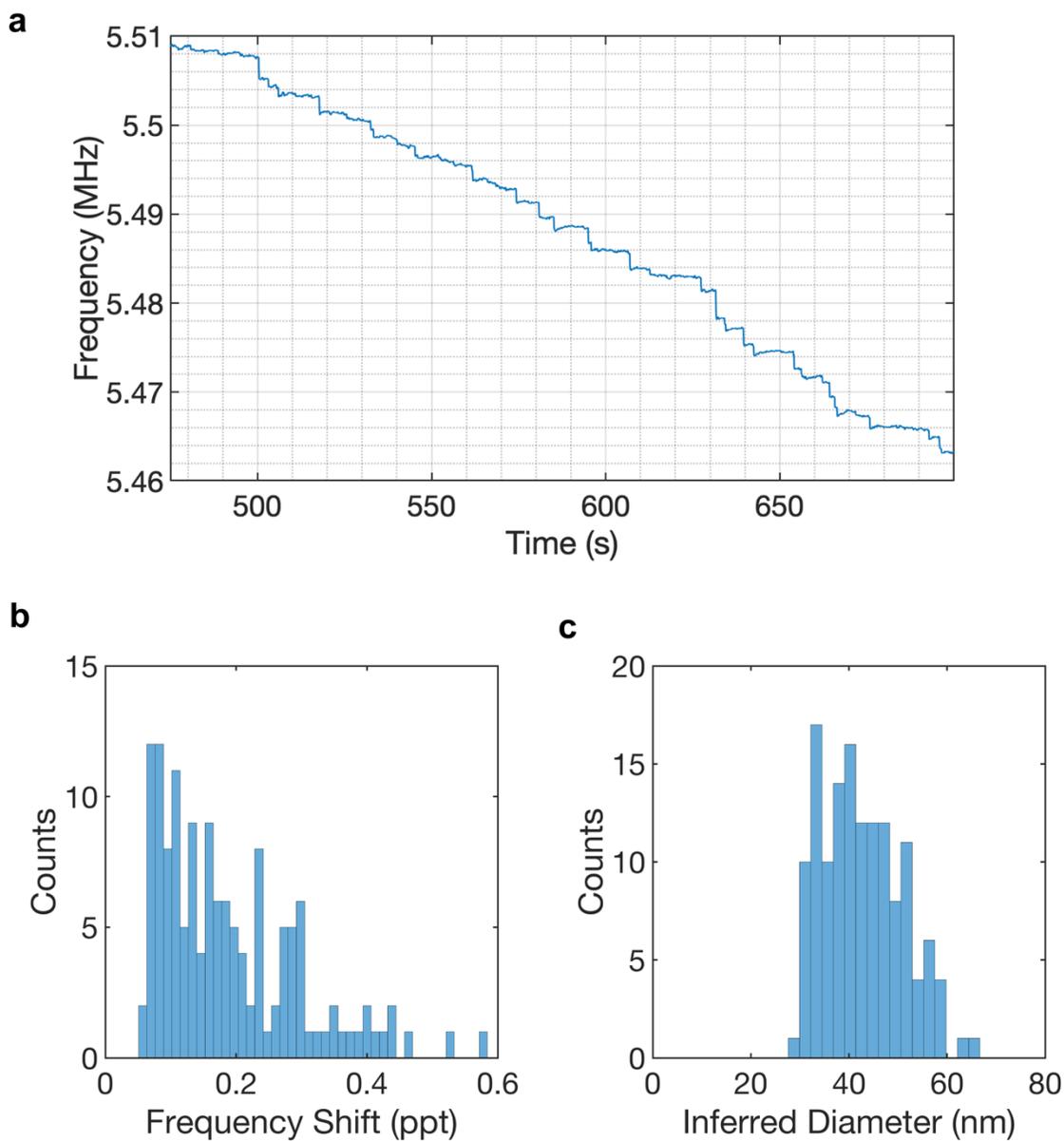

**Figure 4.** 40 nm gold nanoparticle measurements using paddle NEMS device. (a) PLL frequency tracking while spraying the 40 nm gold nanoparticles. (b) Frequency shift histogram and (c) Inferred diameter histogram, which was calculated based on the calculated mass values from the frequency shifts using the density of gold (19.27 g/cm$^3$).

The frequency shifts were converted into mass values and inferred diameter values using the density of gold (19.27 $g/cm^3$) in Figure 4c where events fall around 40 nm. The average value for the diameter calculation yielded 42.6 nm, which is close to the nominal value of 40 nm. Therefore, the results show that this device successfully detected 40 nm GNPs that landed on the middle platform. However, a portion of the events mostly corresponds to lower diameter values due to the particles landing on the supporting beams. This situation can be improved by developing a new polymeric lens architecture that can tightly focus the particles only towards the central platform section, and direct them away from the supporting beams.

The peaks in the inferred diameter histograms of the 200 nm F-PSNP experiments (Table 1 and Figure 3c), as well as GNP experiments (Figure 4c) indicate that the paddle NEMS devices can characterize single-component nanoparticle samples using only one mechanical mode. Also, the calculations are now mode-shape independent, *i.e.* change in mass is directly proportional to the frequency shift. Therefore, the alteration in displacement in the atmosphere can be omitted for the particles landing on the platform owing to the uniformity of mode shape independence. The events landing on the supporting beams constitute a limitation for these devices, however for the analysis of samples composed of single species, mean values of the mass spectrum can be obtained.

**Conclusion**

The paddle NEMS architecture allowed us to have a large region with a uniform displacement profile on the NEMS device. Therefore, this uniformity allowed us to bypass the corrections required due to the alteration in the displacement profiles due to various effects including the attenuation of mechanical waves under atmospheric conditions. We have shown this uniformity in the mode shape by spraying 200 nm F-PSNP to the paddle NEMS device and correlated these measurements with microscopy images. In the mode shape validation experiments, we have shown that the particles landing on the platform created similar frequency shifts. The landing events on the supporting beams created lower frequency shifts due to the smaller displacement values on their landing positions. Moreover, the presence of central platform in paddle NEMS devices increased the capture efficiency by a factor of four. Finally, we performed a 40 nm GNP sensing experiment while PLL tracking where a mean diameter of 42.6 nm was obtained from diameter histogram. It is important to note that we observed landing events on the supporting beams in each experiment that resulted in inaccurate measurement results due to the lower displacement on the supporting beams; this issue can be solved by the implementation of alternative lensing structures to further focus the particles on the central platform or by geometrical optimization of the paddle NEMS device *e.g.* by reducing the supporting beam width. The paddle NEMS architecture provides a practical solution to the alteration in the displacement profiles issue that affects atmospheric pressure NEMS-MS by directly relating the frequency shifts to the particle mass.

**Methods/Experimental**

**Micro and Nanofabrication of Paddle NEMS Devices.** The fabrication flow of the paddle NEMS devices starts with an EBL step using Poly (methyl methacrylate) (PMMA) on the 100 nm thick stoichiometric silicon nitride on a 500 μm thick silicon substrate (University Wafer 1917). The actuation and detection electrodes, alignment markers for the next steps, and contact pads are exposed and developed in this step. After the development, the chip was coated with a 5 nm chrome adhesion layer and a 70 nm gold layer using thermal evaporation. After the first step, one more EBL step takes place to define the paddle NEMS devices including the central platform and the connecting bridges. After the second EBL step and PMMA development, the chip was coated with 40 nm copper as a hard mask to protect the mechanical structures in the dry etching step. Next, the chip underwent two dry etching steps: the first step was the anisotropic etching of silicon nitride, and the second step was the isotropic etching of silicon. Isotropic etching of silicon suspended the paddle NEMS devices. After suspending the devices, the copper masking layer was wet-etched.

Lastly, the polymeric lens structure was implemented by photolithography. First, the device was coated using a thick photoresist (AZ 4533) to a thickness of 3.5 μm and then exposed to UV light through a copper mask to create openings in the active area and contact pads. After the exposure, the chip was developed and then wire bonded to a PCB to be used in the experiments.

**Electrospray Ionization.** The chip was placed in front of an ESI setup in order to facilitate the analyte delivery to the active area of the paddle NEMS device. The solution containing the nanoparticles was placed inside a glass syringe to deliver the solution to the online ESI tip. The ESI flow was then supplied with a high voltage (5.5 kV) to facilitate the ESI process. An extractor lens was placed after the ESI and was also kept at a high voltage (1.3 kV). The details of the ESI setup can be found in the reference.[16]

**Gold Nanoparticle Sample Preparation.** The solution that contains 40 nm nominal diameter particles in DI water was purchased from Nanopartz (A11-40). This stock solution was then diluted using an 80 mM Ammonium Acetate buffer to a final concentration of $8.53 \times 10^9$ particles/mL, including the 10% v/v% methanol to facilitate the electrospray.

**Fluorescent Polystyrene Nanoparticle Sample Preparation.** The aqueous solution that contains the 200 nm fluorescent polystyrene particles was purchased from Thermo Scientific (Fluoro-Max Green Fluorescent Polymer Microspheres, CAT. NO: G200). This solution was diluted to a final concentration of $2.27 \times 10^{10}$ particles/mL using 80 mM Ammonium acetate buffer and 10% v/v% methanol.

ASSOCIATED CONTENT:

**Supporting Information**.

The supporting information contains figures detailing the Allan deviation measurements, and the device modeling such as the calculation of the effective mass.

**Financial Interest Statement:** MSH is the founder of Sensonance Engineering Company; the other authors declare no competing interests.

AUTHOR INFORMATION

Corresponding Author

**Mehmet Selim Hanay** - Department of Mechanical Engineering, and UNAM — Institute of Materials Science and Nanotechnology, Bilkent University, 06800, Ankara, Turkey. selimhanay@bilkent.edu.tr .

**Author Contributions:**

The manuscript was written through contributions of all authors. All authors have given approval to the final version of the manuscript. ‡These authors contributed equally.

B.E.K., M.A. and M.S.H. conceived the idea; B.E.K. and M.A. processed the nanoparticle samples; M.A., B.E.K. and E.K. operated the NEMS-MS system and took the data; C.Y., B.E.K.,

M.A. and E.K. performed nano-fabrication; B.E.K., M.A. and M.S.H. designed the NEMS devices; M.A. and B.E.K. performed the data analysis.

**Funding Sources:** This work was supported by the Scientific and Technological Research Council of Turkey (TÜBİTAK), Grant No. EEEAG-119E503. This project has received funding from the European Research Council (ERC) under the European Union's Horizon 2020 research and innovation programme (grant agreement No 758769).

ACKNOWLEDGMENT

MSH acknowledges fellowship supports from TÜBA and The Science Academy, Turkey. The authors thank Sabancı University SUNUM for nanofabrication support. The authors thank METU MEMS Center for the support with wire bonding. The authors thank Sayedus Salehin, Alper Demir, Ramazan Tufan Erdogan, Hashim Alhmoud, Kadir Ulak and Yegan Erdem for useful discussions.